\input{aipcheck}

\documentclass[
    ,final            
  ]
  {aipproc}

\layoutstyle{8x11double}

\usepackage{multirow}
\usepackage{wasysym}

\begin{document}

\title{Beta Beams Implementation at CERN}

\classification{}
\keywords      {Beta Beams, Decay Ring, CERN, Ion Production, Acceleration, 
  Collective Effects, {\bf EUROnu-WP4-030} }

\author{C. Hansen\footnote{Christian.Hansen@cern.ch}}{
  address={CERN, Geneva, Switzerland}
}

\begin{abstract}
Beta Beam, the concept of generating a pure and intense (anti) neutrino beam by letting accelerated radioactive ions beta decay 
in a storage ring, called Decay Ring (DR), is the base of one of the proposed next generation neutrino oscillation facilities, 
necessary for a complete study of the neutrino oscillation parameter space.
Sensitivities of the unknown neutrino oscillation parameters depend on the Decay Ring's ion intensity 
and of it's duty factor (the filled ratio of the ring).
Therefore efficient ion production, stripping, bunching, acceleration and storing 
are crucial sub-projects under study and development within the Beta Beam collaboration.
Specifically the feasibility of these tasks as parts of a Beta Beam implementation at CERN will be discussed in this report. 
The positive impact of the large $\theta_{13}$ indications from T2K on the Beta Beam performance will also be discussed.
\end{abstract}

\maketitle


\section{Introduction}
\label{sec:intro}

\begin{figure}[ht]
\includegraphics[angle=0, scale= .285]{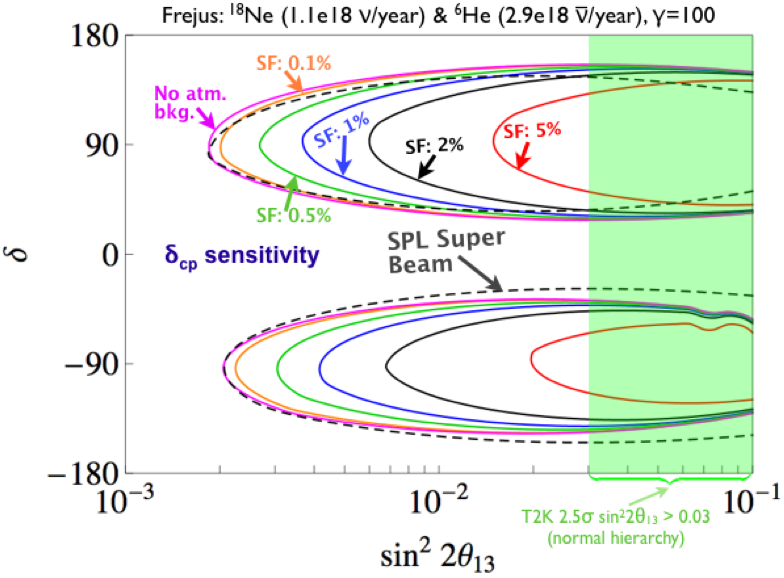}
\caption{The $\theta_{13}$ (left) and  $\delta_{CP}$ (right)
         sensitivities for different suppression factors (SF)~\cite{FernandezMartinez:2009hb}.  
         A SF of 0.01\% (green) gives almost the same result as the case of no atmospheric background (orange). The 0.1\% SF
         (blue) gives similar sensitivity and with 1\% SF (red) the sensitivity has decreased slightly. }
\label{fig:sens}
\end{figure}

The discovery of neutrino oscillations has confirmed that neutrinos are massive and that
their flavor ($\nu_e$, $\nu_{\mu}$, $\nu_{\tau}$) and mass eigenstates ($\nu_1$, $\nu_2$, $\nu_3$) are mixed.
The first oscillation evidence of atmospheric muon neutrinos into tau neutrinos was made by Super-Kamiokande in 
1998 \cite{SK-OSCILLATION}.
Neutrino physics is now in an era of precision measurements of the 
parameters that govern these oscillations:
two $\Delta m^2_{ij} \equiv m_{\nu_i}^2 - m_{\nu_j}^2$ parameters 
($|\Delta m^2_{31}|$, $|\Delta m^2_{21}|$), three mixing angles ($\theta_{23}$, $\theta_{12}$, $\theta_{13}$) 
and a CP violating phase ($\delta_{CP}$). The oscillation phenomena naturally fall into two domains: 
atmospheric, $\nu_{\mu} \to \nu_\tau$, and solar neutrino oscillations, $\nu_e \to \nu_{\mu, \tau}$.
In the solar sector the latest 1$\sigma$ best fits are\footnote{For these values 
normal hierarchy, sign$(\Delta m_{31}^2) > 0$, is assumed. See~\cite{Schwetz:2011zk} for a complete set of values.} 
$\Delta m_{21}^2 = \left(7.59_{-0.18}^{+0.20}\right)\times10^{-5}$~eV$^2$ 
and $\theta_{12} = \left(34.0^{+1.04}_{-0.93}\right) ^{\circ}$ 
and in the atmospheric sector $| \Delta m_{31}^2| = \left(2.50^{+0.09}_{-0.16}\right) \times 10^{-3}$~eV$^2$ 
and $\theta_{23} = \left(46.1^{+3.46}_{-4.02} \right)^{\circ}$~\cite{Schwetz:2011zk}.
According to recent indications from the T2K experiment $\theta_{13} > 5^{\circ}$ with 2.5$\sigma$~\cite{Abe:2011sj}.
A global fit with the latest MINOS results suggests that 1.8$^{\circ} < \theta_{13} < 10.8^{\circ}$ with 3$\sigma$~\cite{Schwetz:2011zk}. 
If it will be proven that $\theta_{13} > 0$ it opens the possibility of discover  $\delta_{CP} \neq 0^{\circ}$ and $\neq 180^{\circ}$
which would mean the existence of CP violation in the leptonic sector. 
Another important unknown value is the sign of $\Delta m_{31}^2$. 
A complete study of all neutrino parameters requires a better characterized neutrino beam with higher flux
than ever available before. 
One of the three present options for the next generation neutrino oscillation facility \cite{FP7-EURONU} 
is the Beta Beam concept \cite{ZUCCHELLI-BETABEAM} wherein it is proposed to store
high energy ($\gamma$~=~100) radioactive ions in a racetrack shaped storage ring with a straight section pointing to a
neutrino detector. 
\begin{figure}[ht]
$\begin{array}{c}
\multicolumn{1}{l}{\mbox{\bf (a)}}             \\
\includegraphics[angle=0, scale= .27]{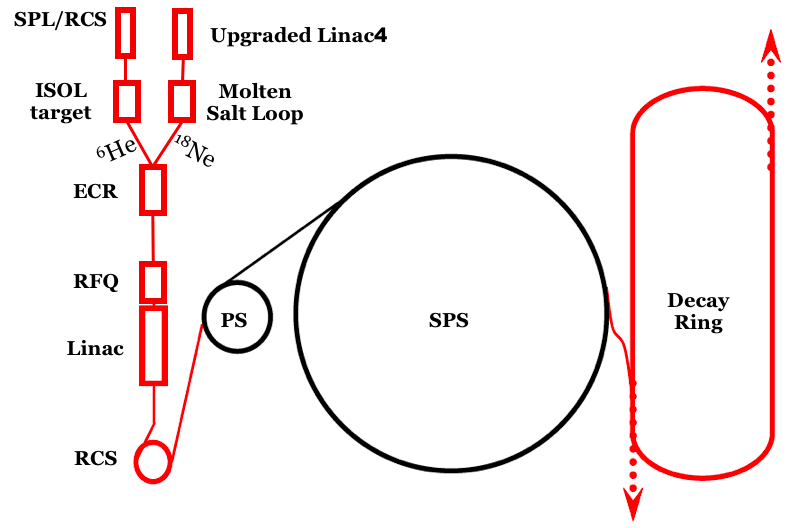}  \\
\multicolumn{1}{l}{\mbox{\bf (b)}} \\
\includegraphics[angle=0, scale= .28]{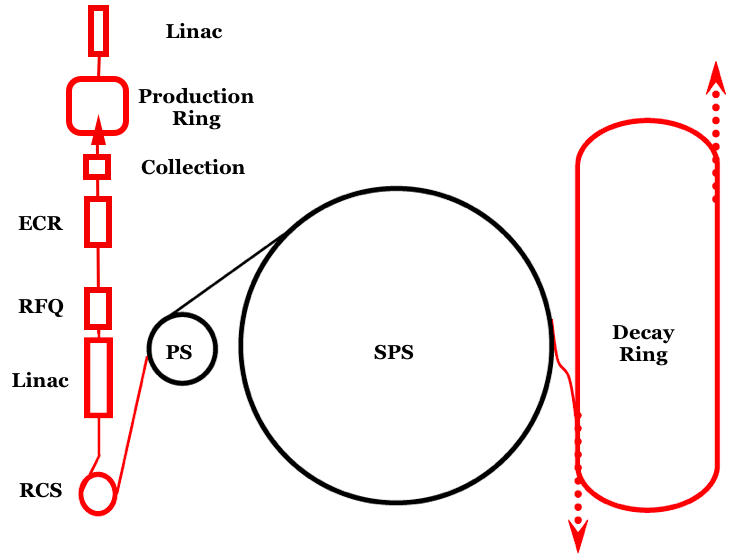} \\ 
\end{array}$
\caption{Beta Beam baseline implementation at CERN for low Q ions ($^6$He and $^{18}$Ne) {\bf (a)} and
         high Q ions ($^8$Li and $^8$B) {\bf (b)}.}
\label{fig:baselines}
\end{figure}
The ions have to have optimized decay time, to maximize the neutrino flux, and Q-value\footnote{The ion's reaction energy. It defines the neutrino energy.}, for the chosen baseline, 
and to produce both a neutrino and anti-neutrino beam a pair of $\beta^-$ and $\beta^+$ decaying ions have to be used. 
Ions that beta-decay in the straight section emit electron (anti) neutrinos in a pure 
(almost only $\bar{\nu}_e$ or $\nu_e$) beam with an opening angle of $1/\gamma$.
The aimed annual (anti) neutrino flux of (2.9e18) 1.1e18 from ($\beta^-$) $\beta^+$ decaying ($^6$He) $^{18}$Ne ions 
gives $\delta_{CP}$ sensitivities shown in fig.~\ref{fig:sens}.
The distance to the first $\nu_e \to \nu_{\mu}$-oscillation maxima for this ion-pair (with Q-values around 3~MeV) 
and thereby to the location of the detector is about 130~km. 
Corresponding sensitivity studies for an ion-pair with higher Q-values (around 13~MeV), $^8$Li and $^8$B, have been made
and show that about a factor 5 times higher neutrino flux is necessary for similar sensitivities~\cite{FernandezMartinez:2009hb}. 
The baseline for this case is around 700~km. 
For both the low and high Q-value ion pairs the baselines studied in this report start at CERN and use PS and SPS as part of the accelerating scheme (see fig.~\ref{fig:baselines}). 
The sensitivities depend also on the suppression factor (SF) of the detector, which is the same as the duty factor for the 
storage ring. This storage ring is called the ``Decay Ring'' (DR) and has the same circumference as SPS, $C$~=~6911.6~m. 
The required SF and its impact will be discussed in this report after ion production, preparation, bunching and 
acceleration is presented.

\section{Ion Production}
\label{sec:production}

\begin{figure}[ht]
\includegraphics[angle=0, scale= .285]{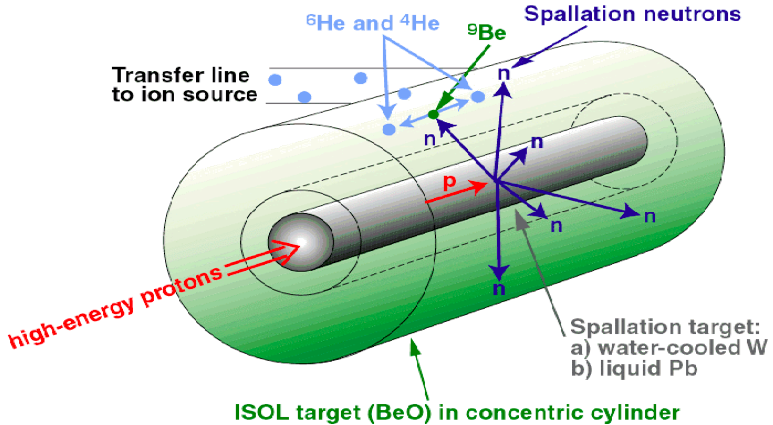}
\caption{Spallation target (W or Pb) and ISOL target (BeO) for $^6$He production.}
\label{fig:spallation}
\end{figure}

The CERN implementation of the baseline for the ion pair with low Q-value, $^6$He and $^{18}$Ne, 
will consist of two different production schemes depending on the ion. 
For the $^6$He ion production the ISOLDE~\cite{Kugler:1993vm} method will be used.
High energy protons will bombard a heavy spallation target (either tungsten or lead) so that spallation
neutrons are forced out to and react with the surrounding ISOL target (normally BeO) so that
$^6$He ions are produced (see figure~\ref{fig:spallation}). 
This is a standard method and the production rate can be estimated by
scaling and would depend only on the proton source. Assuming a 200~kW Super Proton Linac (SPL) that gives
2~GeV protons or a 20~kW Rapid Cycling Synchrotron (RCS)\footnote{This RCS are under consideration
for the LHC upgrade and Hie-ISOLDe~\cite{DElia:2009xb} would also benefit from it.}, 
also giving 2~GeV protons, the production rate would be 5$\times10^{13}$ $^6$He/s.

\begin{figure}[ht]
\includegraphics[angle=0, scale= .25]{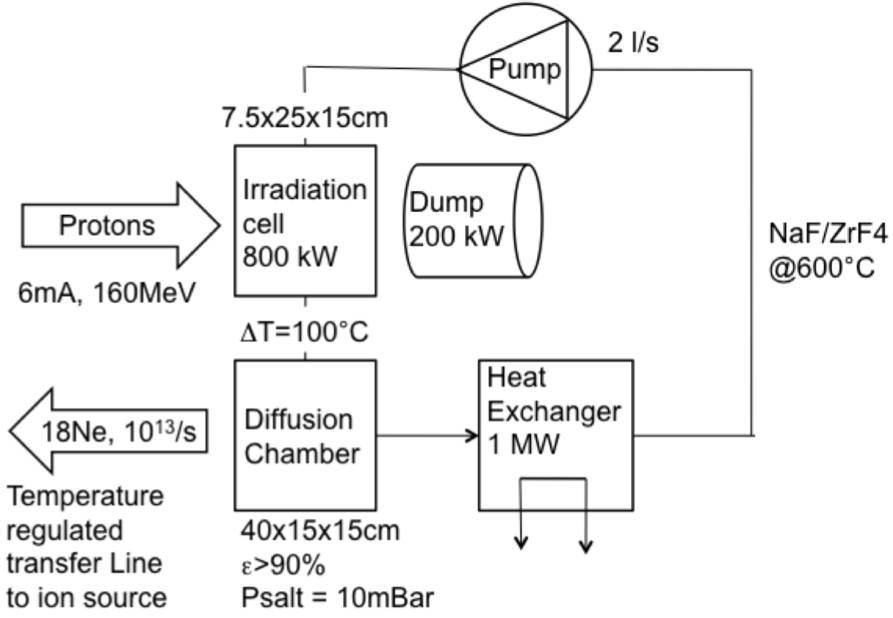}
\caption{The molten salt loop to produce $^{18}$Ne with two reactions; 
  $^{19}F(p,2n)^{18}Ne$ and $^{23}Na(p,X)^{18}Ne$. If an upgrade of CERN's Linac4 gives
        160~MeV protons this target station could give 10$^{13}$~$^{18}$Ne/s~\cite{MENDONCA_NUFACT11}.}
\label{fig:molten}
\end{figure}

The $^{18}$Ne production would need 160~MeV protons from a linac, possibly from an upgraded Linac4, 
to hit a molten salt (NaF) loop. Two reactions would then contribute
to the $^{18}$Ne production; $^{19}F(p,2n)^{18}Ne$ and $^{23}Na(p,X)^{18}Ne$ 
(see figure~\ref{fig:molten}).
According to studies 1~MW target station (only twice as powerful as the current operating target station 
for CNGS) would provide about 10$^{13}$~$^{18}$Ne/s.
This is however a novel idea and a proof of principle experiment is scheduled for November at CERN~\cite{MENDONCA_NUFACT11}. 

\begin{figure}[ht]
\includegraphics[angle=0, scale= .36]{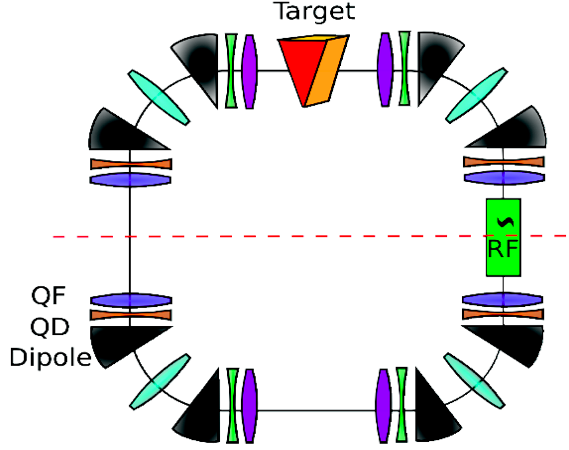}
\caption{Production ring for $^8$Li and $^8$B production~\cite{Schaumann:2009zz}.}
\label{fig:productionring}
\end{figure}

\begin{figure}[ht]
\includegraphics[angle=0, scale= .23]{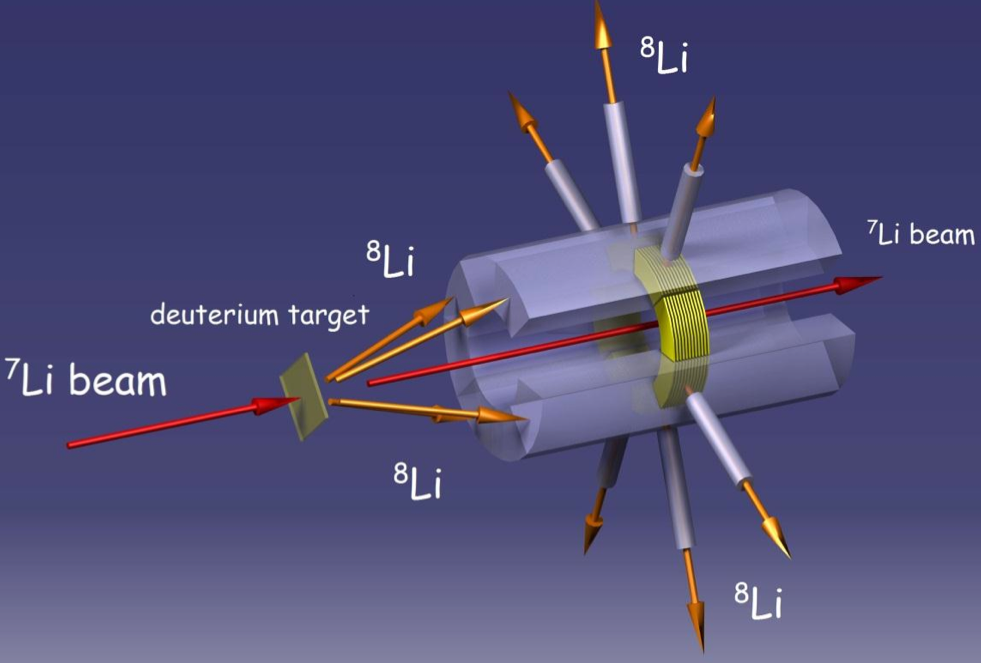}
\caption{Collection device with thin, 2~$\mu$m, catcher foils~\cite{SEMEN_NUFACT11}.}
\label{fig:collection}
\end{figure}

Another novel concept is the production of $^8$Li and $^8$B ions with the so called 
``Production Ring'' (PR)~\cite{RUBBIA-BEAMCOOLING}. For the $^8$Li production 25~MeV $^7$Li 
ions enters the PR that is a small storage ring with a wedge shaped gas-jet deuterium 
target (see figure~\ref{fig:productionring}). 
Multiple passes through this target cause ion-cooling, stripping and production 
through the reaction $^2_1$H+$^7_3$Li$\to^8_3$Li+$p$. The produced $^8$Li ions would then
exit the target with an angular distribution and be collected by 2~$\mu$m thick catcher foils
and then diffused away with a cold finger (see figure~\ref{fig:collection}). 
The efficiency of this collection device are to be determined in ongoing studies~\cite{SEMEN_NUFACT11}~.
The same concept would be used for $^8$B production but with the reaction $^3_2$He+$^6_3$Li$\to^8_5$B+$n$.
The cross-section and angular distribution of this reaction is however unknown. 
Therefore measurements have been performed and analysis is ongoing~\cite{VLADIMIR_NUFACT11}. 
Production Ring studies have shown that there are difficulties to reach the required target 
density~\cite{Benedetto:2010zza} therefore new ideas that suggests direct kinematics through a $^7$Li 
film target will be tested~\cite{NOLEN_NUFACT11}.
Further details about ion production for the Beta Beams will be shown in~\cite{STORA_NUFACT11}.

\section{Beam Preparation}
\label{sec:ecr}

The gas of the radioactive atoms from the ion production (see previous section) 
diffuse into a volume with a magnetic field, $B$. This field makes free cold electrons
move in a circular motion with the cyclotron angular frequency $\omega_{ce} = eB/m$ where
$e$ is the elementary charge and $m$ is the mass of the electron. 
An electro magnetic wave with the same high frequency, $\omega_{hf} = \omega_{ce}$, is applied
and allow Electron Cyclotron Resonance (ECR) heating. 
Due to their subsequent energy, when they collide the atoms of the gas, they are able to ionize them. 
This ECR ion source has best efficiency producing singly charged
ions so a stripper giving fully stripped ions is assumed after the ECR. 
\begin{figure}[ht]
\includegraphics[angle=0, scale= .23]{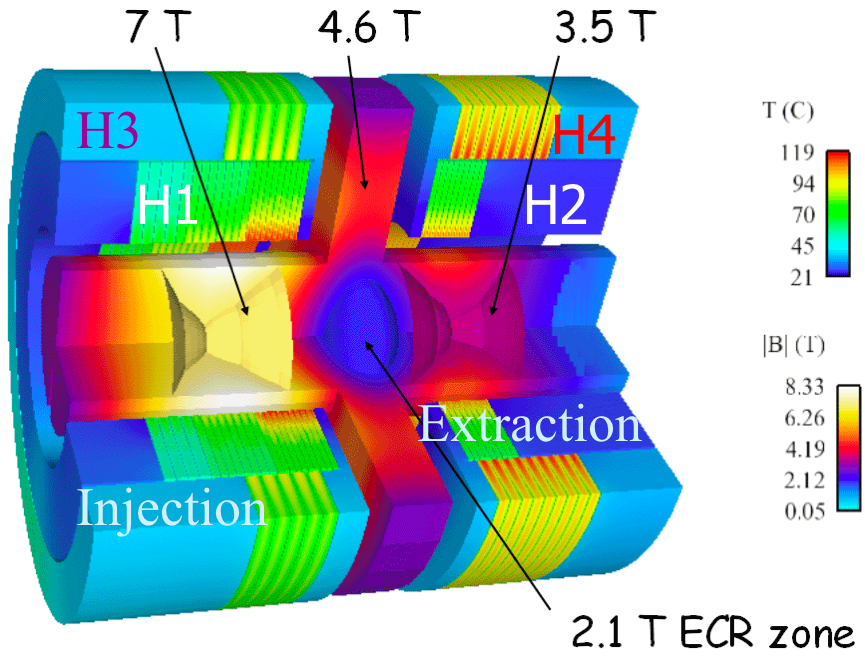}
\caption{Magnetic field and temperature map of the 60~GHz ECR source~\cite{LAMY_NUFACT11}.}
\label{fig:ecr}
\end{figure}
R\&D is ongoing for a 60~GHz ECR source (see figure~\ref{fig:ecr}), which would be the most powerful ECR source to
date~\cite{LAMY_NUFACT11}. It operates in pulse mode so that 50~$\mu$s pulses
of monocharged ions will be extracted from the ECR.
The estimated efficiencies for the Beta Beam ions in  
this case are 30\% for $^6$He, 20\% for $^{18}$Ne and 3 to 10 times less for $^8$Li and $^8$B.

\section{Bunching and Acceleration}
\label{sec:acc}

\begin{figure}[ht]
\includegraphics[angle=0, scale= .23]{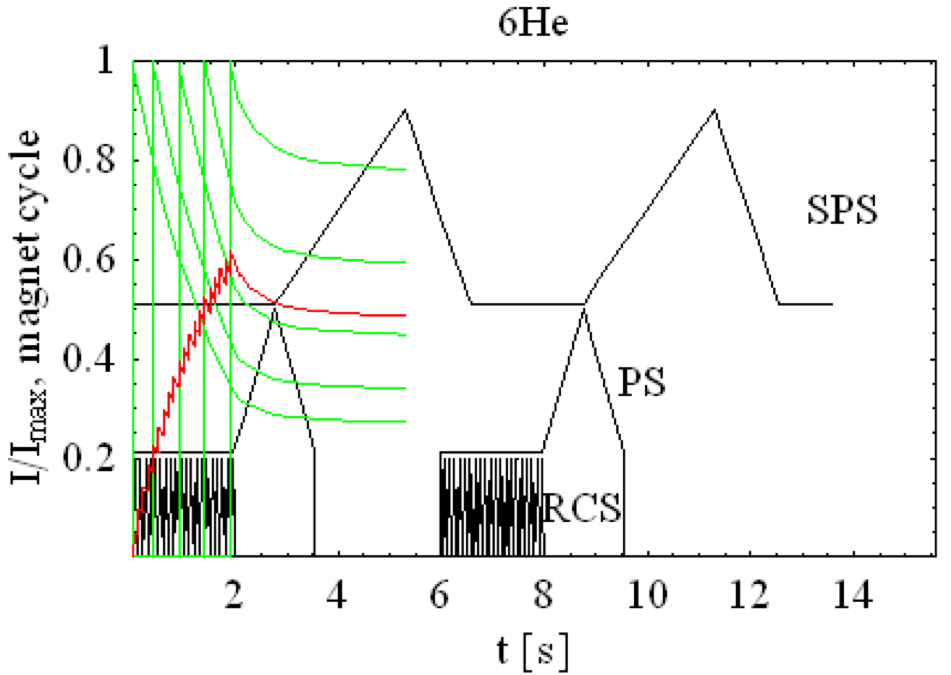}
\caption{Single bunch intensities are indicated in green for the
1st, 5th, 10th and 20th bunch and the total number of helium ions
are indicted in red. The schematic time structure of the RCS, PS and SPS magnet cycles 
are also shown in black~\cite{Benedikt:2006zk}.}
\label{fig:psaccum}
\end{figure}

Fully stripped ions enter the  Beta Beam acceleration
complex in 50~$\mu$s dc bunches and with about 8~keV/u.
Prebunching and preacceleration is done with a Radio Frequency Quadrupole (RFQ)
and a Linac. A moderate acceleration gradient of 3-6~MV/m is assumed so
normal RF cavities can be used. The Linac will accelerate the ions up to
100~MeV/u before they are transfered to a Rapid Cycling Synchrotron (RCS).
The first task for the RCS is to create bunches with high ion intensity.
This is done by multiturn injection. With the current design of the RCS 
for the Beta Beams~\cite{Lachaize:2008zz} the revolution time is 1.96~$\mu$s. 
The ECR pulses are 50~$\mu$s so the injection takes 26 turns. 
Thanks to phase-space rotation the 26 injections fit into one single RCS bunch 
with 50~\% efficiency.
The second task of the RCS is  acceleration up to about 15~Tm before the
high intensity bunch is transfered to CERN's Proton Synchrotron (PS).
The PS collects 20 RCS bunches one by one. It will be 1.9 seconds between 
the first and last injection. There will therefore be intensity differences in the 
bunches when all 20 bunches are injected to the PS due to radioactive decay
(see figure~\ref{fig:psaccum}). 
After injection the batch of 20 bunches are accelerated up to maximum PS 
capacity, 86.7~Tm, and then transfered to CERN's Super Proton Synchrotron (SPS). 
For proton operation the PS bunches are normally split before entering 
SPS so that they can be handled by the 200~MHz (h = 4620) SPS RF system.
However, for the Beta Beams we assume that an additional slower, 40~MHz (h = 924), 
RF system has been implemented into the SPS.
This is to mitigate space charge effects. 
Beta Beams have a high ion intensity and specially a high charge intensity. 
For example for nominal Beta Beam neutrino fluxes we need 4.9$\times$10$^{12}$ 
charges per bunch in the SPS and the comparable most intense beam accelerated
in the SPS is 3.5$\times$10$^{11}$ charges per bunch. 
Space charge effects are worse for low energies 
(it goes as $\gamma^{-2}$\cite{Ruggio:1995}).
Therefore large bunches (same size as given by PS) are taken care of by the 
slow 40~MHz and 1 MV RF system at SPS injection where $\gamma$~=~9. 
\begin{figure}[ht]
\includegraphics[angle=0, scale= .2]{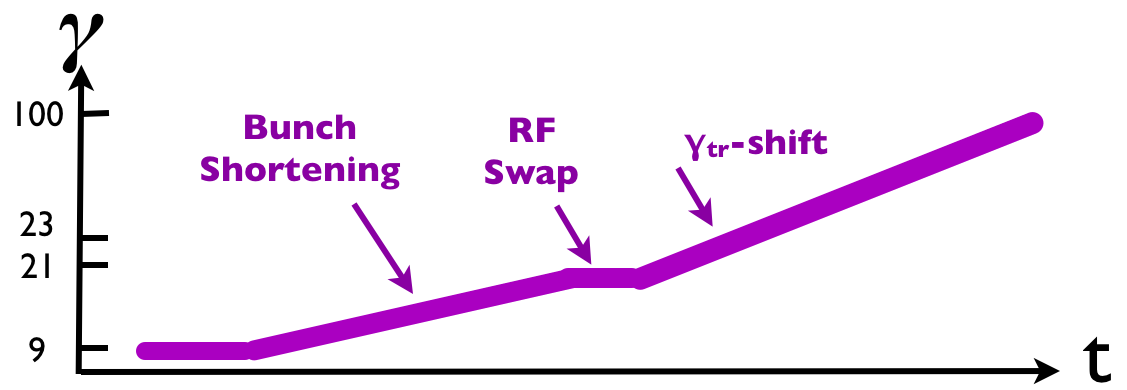}
\caption{A simplified diagram showing how one acceleration part of the  
         SPS RF cycle could look like.}
\label{fig:spsRF}
\end{figure}
After initial acceleration space charge effect is less crucial and the bunches
are shortened so that the RF can swap to the normal 200~MHz and 8~MV RF system. 
With this RF system all ions can then be accelerated across transition 
and then up to a $\gamma$~=~100, for all ions.

\section{Accelerated Intensities}
\label{sec:intens}

The ion intensities through the whole Beta Beam complex are calculated~\cite{WILDNER-MATHEMATICA} starting with the 
source rate and taking into account the radioactive decay and the efficiencies of 
all components (discussed above). Results are shown in table~\ref{tab:intens}.
They are shown for the low-Q ion pair only since source-rate and efficiencies are still too 
uncertain for the high-Q ion pair. 
Intensities are shown per bunch at the end of each components cycle, i.e. the last row shows the 
intensities in SPS when the ions have reached $\gamma$~=~100. 
The injection into the DR and the DR intensities will be discussed in the next section. 
\begin{table}[htbp]
\begin{tabular}{l|cc}
                             & \ \ \ \ \ \ \ \ \    $^{18}$Ne \ \ \ \ \ \ \ \ \    &\ \ \ \ \ \ \ \ \     $^{6}$He \ \ \ \ \ \ \ \ \       \\
\hline
Source rate [10$^{13}$/s]     &  1.2     &  5.0  \\
ECR [10$^{11}$/pulse]         &  2.4     &  14.4  \\
RCS [10$^{11}$/bunch]         &  1.2     & 7.0   \\
PS  [10$^{11}$/bunch]         &  1.0     & 4.0   \\
SPS [10$^{11}$/bunch]         &  1.0     & 3.8   \\
\end{tabular}
\caption{Ion bunch intensities after each stage in the Beta Beam complex before the Decay Ring.
Corresponding values for $^8$B and $^8$Li are not shown here due to large uncertainty 
in source rate and ECR efficiency. 
}
\label{tab:intens}
\end{table}

\section{Decay Ring}
\label{sec:dr}

The Decay Ring (DR) is a racetrack shaped storage ring aiming the neutrino beam from the 
decaying ions to a neutrino detector in a big cave somewhere in Europe.
For the low Q ions the DR will have a 0.6$^{\circ}$ declination angle pointing 
towards the Frejus tunnel (French/Italian border) and for the low Q ions the DR will decline with a 
2.9$^{\circ}$ or 3.3$^{\circ}$ angle pointing towards Canfranc (Spain) or Gran Sasso (Italy) respectively. 
In the current design~\cite{ANTOINE_NUFACT11} the DR has the same circumference as SPS, 6911.6~m, and the 
efficient straight section is 39\%. This gives bending radii as small as 121~m 
so that superconducting dipole magnets, giving a 4 to 8~T magnetic field, are necessary.
One important aspect of the DR is that the ion bunches can only fill a small part 
of it's total length.  
Before the hints of large $\theta_{13}$ (see next section) it was assumed a duty factor as small 
as 0.58\% is necessary for physics reach. 
The 20 bunches then had to be limited to a bunch length of 6911.6m$\times$0.58\%/20 = 2~m each.
There is also a strong requirement of ion intensities in the DR.
With 20 bunches, the number $^{18}$Ne ($^6$He) per bunch have to be $3.4\times10^{12}$ ($4.5\times10^{12}$) 
(and about 5 times more of $^8$B ($^8$Li) ions) to reach the nominal (anti) neutrino fluxes. 

\subsection{Decay Ring Injection}
\label{sec:dr_inj}

\begin{figure}[ht]
$\begin{array}{c}
\multicolumn{1}{l}{\mbox{\bf (a)}}       \\ [0.1cm]
\includegraphics[angle=0, scale= .33]{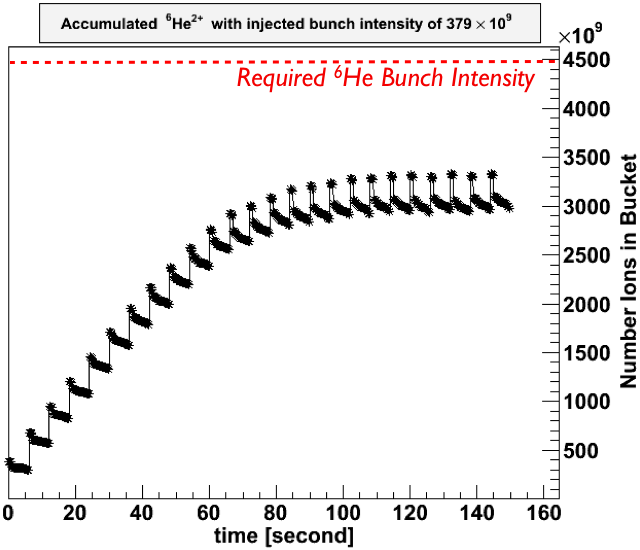} \\ [0.1cm]
\multicolumn{1}{l}{\mbox{\bf (b)}}       \\ [0.1cm]
\includegraphics[angle=0, scale= .32]{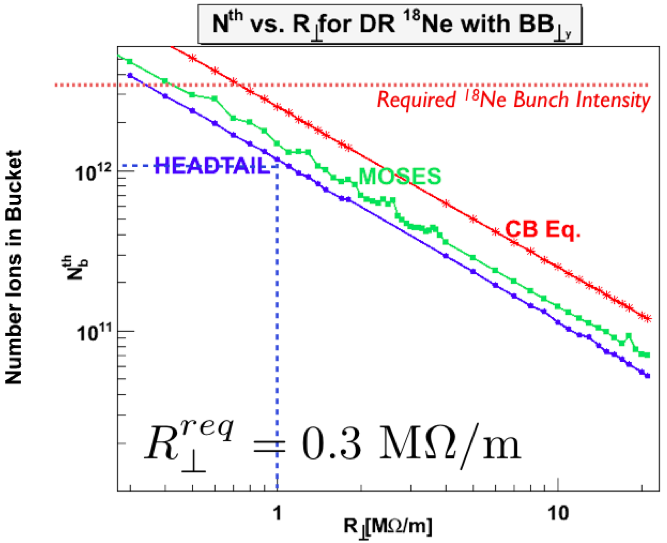} \\ [0.1cm]
\end{array}$
\caption{{\bf (a)}Accumulation of $^6$He ions in a DR bunch. Saturation occurs at 3$\times$10$^{12}$ $^6$He/bunch,
  less than the required 4.5$\times$10$^{12}$ $^6$He/bunch. {\bf (b)} Bunch intensity limits according to HEADTAIL~\cite{headtail}, 
  MOSES~\cite{Chin:1988xj} and the Coasting Beam equation~\cite{Metral:2004vi} for $^{18}$Ne in the DR.}
\label{fig:smallSF}
\end{figure}

A dedicated injection scheme of the SPS bunches into the DR has been developed~\cite{HANCOCK-STACKING, HEINRICH} 
in order to increase the ion bunch intensities and limit the bunch lengths. This scheme uses two RF systems and 
includes catching the SPS bunch injected off-momentum, merging the fresh SPS bunch into the 
existing DR bunch (to increase the bunch intensity) and collimation at $dp/p$~=~2.5$\permil$ 
(to limit the bunch length to 2~m).
Numbers of ions per bunch injected from SPS are shown in the last row in table~\ref{tab:intens}.
During the first injections only ions that are not captured are collimated away (capture efficiency is 
about 90\% for all ions \cite{HEINRICH}). Due to RF-gymnastics during merging the phase-space 
of the DR bunch increases and after about 15 merges many ions are collimated after each injection so that 
the bunch length do not exceeds 2~m. Due to this collimation and to the constant decay of the radioactive 
ions the number of ions per bunch saturates after about 20 injections. For the $^6$He case where 
3.8$\times$10$^{11}$ ions are injected per bunch the accumulation of the DR bunch saturates at about
$N_{b_{sat}}$~=~3$\times$10$^{12}$ $^6$He/bunch i.e. below the required 4.5$\times$10$^{12}$ $^6$He/bunch 
(see figure~\ref{fig:smallSF}{\bf (a)}).
From the number saturated ions per bunch we get the annual neutrino flux from
\begin{equation}
\phi = \frac{N_{b_{sat}} N_B \ell_{eff} T_{eff}}{T_C} \left( 1 - 2^{-\frac{T_C}{\gamma t_{1/2}}}  \right)  \  ,
\label{eq:phi}
\end{equation} 
where  $T_{eff}$~=~10$^7$~seconds\footnote{An efficient Beta Beam year is assumed to be one third of a year.} 
and all other parameters are given in table~\ref{tab:ION-ACCUM-RESULT} and \ref{tab:values}.
The results from these studies and for SF~=~0.58\% are summarized in first and second column in 
table~\ref{tab:ION-ACCUM-RESULT} for $^{18}$Ne and $^{6}$He respectively.  

\subsection{Decay Ring Storage}
\label{sec:dr_store}

High intensity bunches can have non-negligible amount of charges which could cause
the particles to interact with each other and with the vacuum chamber. These, so called ``Collective Effects'',
could cause many different reasons for beam instability. For example ``single bunch resonance broad band impedance'' 
appears when electro magnetic wake fields from the head of the bunch are captured by cavities in the vacuum pipe and 
interact with the tail of the bunch.  
If the {\it quality factor} is $Q$~=~$R\sqrt{C/L}$ and the {\it resonance frequency} is $\omega_r$~=~$1/\sqrt{LC}$
the resonance impedance can be modelled as an RLC circuit \cite{Chin:8thJoint} in the transverse plane as
\begin{equation}
Z_{\perp}(\omega) = \frac{R_{\perp}\frac{\omega_r}{\omega}}{1+iQ\left(\frac{\omega_r}{\omega}-\frac{\omega}{\omega_r} \right)}
\label{eq:rlc}
\end{equation} 
where the {\it transverse shunt impedance}, $R_{\perp}$, is a value indicating the total divergence from a perfectly 
smooth vacuum pipe around the whole ring. 
The value for RHIC is about $R_{\perp}^{RHIC}$~$\approx$~2~M$\Omega$/m~\cite{Fischer:2008zzc}.
The DR will be a modern machine so we will assume it will have a smooth vacuum pipe design and therefore
twice as good shunt impedance; $R_{\perp}^{DR}$~$\approx$~1~M$\Omega$/m~\cite{rumolo_private}. 
However, simulations, modeled on eq.~(\ref{eq:rlc}), with a tracking program called HEADTAIL~\cite{headtail},
theoretical calculations, with a program called MOSES~\cite{Chin:1988xj}, and estimations with a simplified equation
called Coasting Beam equation (CB eq.)~\cite{Metral:2004vi} all claim that although $R_{\perp}^{DR}$~$\approx$~1~M$\Omega$/m
resonance impedance would cause instable beams below the required $3.4\times10^{12}$~$^{18}$Ne/bunch (see figure~\ref{fig:smallSF}{\bf (b)}). 
For example, according to HEADTAIL the $^{18}$Ne ion beam would be unstable if there were more than 10$^{12}$ ions per bunch, i.e. more than 
a factor 3 below the required.
All parameters used for these simulations and calculations are listed in table~\ref{tab:valuesIonDep} and \ref{tab:values}.

\vspace{1cm}
\begin{table}[htbp]
\begin{tabular}{lc|cc}
\hline
Parameters & {\it Description}  & DR $^{18}$Ne & DR $^6$He \\ 
\hline
$Z$  & {\it Charge Num.}  & 10 & 2 \\ 
$A$ &  {\it Mass Num.}  & 18 & 6 \\ 
$t_{1/2}$ [s] &  {\it Half Life}  & 1.67 & 0.81 \\ 
$V_{RF}$ [MV] & {\it Voltage}   & 26.75 \\
$E_{rest}$ [MeV] & {\it Rest Energy}   & 16767.10 &  5605.54 \\ 
$L_{b}$ [m] & {\it Bunch Length}   &  1.97  & 1.97 \\ 
$\delta_{max}$ [10$^{-3}$] & {\it Mom. Spread} & 2.5 & 2.5  \\ 
\multicolumn{2}{l|}{$\varepsilon_{N_{x}} (1\sigma)$ [$\pi$m$\times$rad]} & 1.48e-05  & 1.48e-05  \\ 
\multicolumn{2}{l|}{$\varepsilon_{N_{y}} (1\sigma)$ [$\pi$m$\times$rad]} & 7.90e-06  & 7.90e-06  \\ 
\hline
\multicolumn{2}{l|}{$E_{tot}$ [GeV] $= \gamma \times E_{rest}$}  &  1676.7  & 560.6  \\ 
\multicolumn{2}{l|}{$\tau_{b}$ [ns] $=\frac{L_{b}}{\beta c}$}   &   6.57 & 6.57 \\ 
\multicolumn{2}{l|}{$\sigma_{\delta}[e^{-3}]=\frac{\delta_{max}}{2}$ } & 1.25  & 1.25  \\
\multicolumn{2}{l|}{$\varepsilon_{l}^{^{2\sigma}}$ [eVs] $=\frac{\pi}{2}\beta^{2}E_{tot}\tau_{b}\delta_{max}$} & 43.20  &  13.95  \\  
\multicolumn{2}{l|}{$Q_{s}$ [10$^{-3}$] =$\sqrt{\frac{hZeV|\eta \cos\phi_{s}|}{2\pi\beta^{2}E_{tot}}}$ }& 8.1  &   3.7 \\ 
\multicolumn{2}{l|}{$r_{0}$ [am] $= r_{p}Z^{2}/A$}  & 8.53 & 1.02  \\ 
\hline
\end{tabular}
\caption{Input parameters (some from~\cite{FP6-FINAL}). Parameters below the line are all calculated. 
}
\label{tab:valuesIonDep}
\end{table}

\begin{table}[htbp]
\begin{tabular}{l|lc}
\hline
Parameters & {\it Description}  & DR \\ 
\hline
$h$ & {\it Harmonic Number}  & 924 \\ 
$C$ [m] & {\it Circumference}  & 6911.6 \\ 
$\ell_{eff}$ & {\it Efficient Straight Section}  & 39$\%$ \\
$\rho$ [m] & {\it Magnetic Radius}  & 155.6 \\ 
$\gamma_{tr}$ & {\it Gamma Transition}  & 18.6 \\ 
$\gamma$ & {\it Relativistic Gamma} & 100.0  \\ 
$Q_{x}$ & {\it Horizontal Tune}  &  21.23  \\ 
$Q_{y}$ & {\it Vertical Tune}  &  17.16  \\ 
$\langle \beta \rangle_{x}$ [m] & {\it Av. x-$\beta$tron Func.}   & 124.70  \\ 
$\langle \beta \rangle_{y}$ [m] & {\it Av. y-$\beta$tron Func.}  & 160.40  \\ 
$\langle D \rangle_{x}$ [m] & {\it Av. Dispersion}  &  -0.60  \\ 
$\xi_{x,y}$ & {\it x,y Chromaticity}  &   0.0  \\ 
$b_{x}$ [cm] & {\it x Pipe Size}  &   16.0  \\ 
$b_{y}$ [cm] & {\it y Pipe Size}  &    16.0  \\ 
\hline
$Q_{\perp}$ & {\it Quality Factor} &    1.0  \\ 
$f_{r}$  [GHz] & {\it Resonance Frequency}   &   1.0  \\ 
$R_{\perp}  [\frac{M\Omega}{m}]$ & {\it Shunt Impedance}   &  1.0  \\ 
\hline
$\beta=\sqrt{1-\gamma^{-2}}$  & {\it Relativistic Beta}  &   1.00  \\ 
$\eta=\gamma_{tr}^{-2}-\gamma^{-2}$ & {\it Phase Slip Factor} & 2.8e-3  \\ 
$T_{rev}$[$\mu$s]$=\frac{C}{\beta c}$ & {\it Revolution Time}   & 23.06  \\
$f_{rev}$[Hz]$=\frac{1}{T_{rev}}$ & {\it Revolution Frequency}  &   0.27e6  \\  
$R$ [m] $= C / 2\pi$ & {\it Machine Radius}  &  1100  \\ 
$\omega_{c}$[GHz]$=\frac{\beta c}{b_{x,y}}$ & {\it Cut-Off Angular Freq.}  &    1.87  \\ 
\hline
\end{tabular}
\caption{Input parameters (some from the previous Beta Beam Decay Ring design report~\cite{FP6-FINAL} and
  some from updated design~\cite{ANTOINE_IPAC11}) above the first line.
  Assumed transversal impedance parameters between the lines. Calculated parameters below the last line. 
  These parameters are the same for the different isotopes.}
\label{tab:values}
\end{table} 

\section{Impact of Large $\theta_{13}$}
\label{sec:largetheta}

In the previous section we showed two reasons why the DR could not use the ions provided by the 
rest of the Beta Beam complex efficiently enough to reach required neutrino flux;
Not enough ions could be accumulated per bunch with the merging technique and the bunch 
intensity limit due to collective effects were also below the required. The underlying 
reason for both of these effects is the limited suppression factor; 0.58\%.  
\begin{figure}[ht]
$\begin{array}{c}
\multicolumn{1}{l}{\mbox{\bf (a)}}       \\ [0.1cm]
\includegraphics[angle=0, scale= .33]{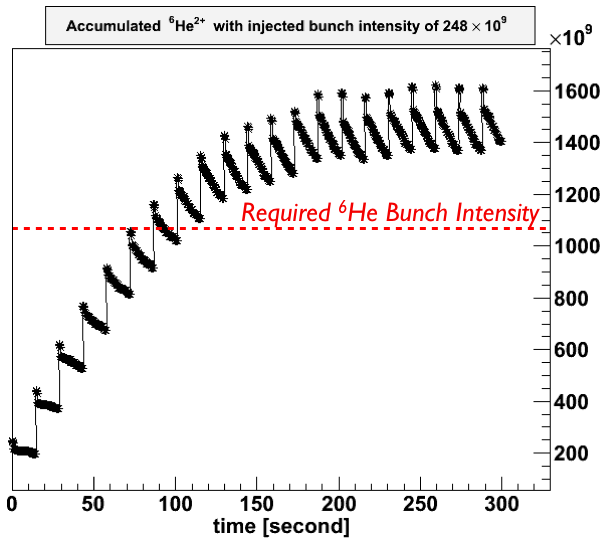} \\ [0.1cm]
\multicolumn{1}{l}{\mbox{\bf (b)}}       \\ [0.1cm]
\includegraphics[angle=0, scale= .32]{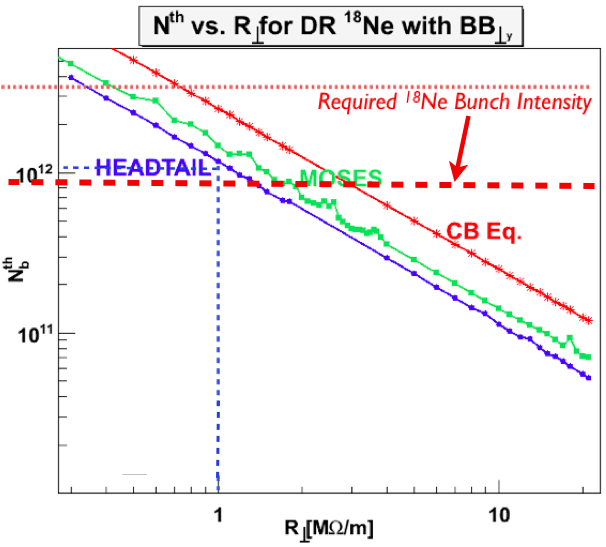} \\ [0.1cm]
\end{array}$
\caption{In case of large $\theta_{13}$ SF can be relaxed to 2.3\% and then $^6$He saturation {\bf (a)} occurs at 
  1.4$\times$10$^{12}$ $^6$He/bunch, more than the required 1.1$\times$10$^{12}$ $^6$He/bunch.  Bunch intensity 
  limits due to collective effects {\bf (b)} also relax above required.}
\label{fig:bigSF}
\end{figure}
New results from the T2K experiment show however indications~\cite{Abe:2011sj} 
that would relax the duty factor requirements in the DR and open new opportunities for the Beta Beams.
Figure~\ref{fig:sens} shows that in the $\theta_{13}$ region allowed by T2K's 6 $\nu_e$ candidate 
events (green zone indicated in fig.~~\ref{fig:sens}) the suppression factor is much less crucial. 
A suppression factor of 2\% (black line) would give more or less as good $\delta_{CP}$ sensitivity 
as SF~=~0.5\%. The impact of filling four times more of the DR, 2.3\% instead of 0.58\%, has therefore
been studied. This opportunity could be used in different ways but so far only one has been studied; 
to fill the DR with the same bunches, 2~m long, but with 4 times as many bunches. 
This simple solution can be achieved by letting SPS wait for 4 PS batches before the SPS 
batch of 80 bunches is injected into the DR. More ions will then have decayed and average SPS bunch intensity 
will be lower; 0.9$\times$10$^{11}$~$^{18}$Ne/bunch and  2.5$\times$10$^{11}$~$^{6}$He/bunch.
The average DR bunch intensity will therefore saturate at a lower value (compare figure~\ref{fig:bigSF}{\bf (a)}
with figure~\ref{fig:smallSF}{\bf (a)}). The required bunch intensity has however decreased with 
a factor 4 (since we have 4 times more bunches) so that the accumulation succeeds to provide with enough neutrino flux.
The result for these studies and for SF~=~2.3\% are summarized in third (for $^{18}$Ne) and fourth (for $^6$He) 
column in table~\ref{tab:ION-ACCUM-RESULT} where also comparison with the nominal neutrino fluxes are made.
 
Thanks to the decrease of required bunch intensity the 80 ion bunches can also store enough
ions before resonance impedance cause beam instability (see figure~\ref{fig:bigSF}{\bf (b)}). 
However other collective effects reasons will need to be studied to confirm this.

\begin{table}[htbp]
\begin{tabular}{l|cccc}
\hline
Suppression Factor, SF                                           & \multicolumn{2}{c}{0.58\%}  & \multicolumn{2}{c}{2.3\%} \\
Number Bunches, $N_B$                                           & \multicolumn{2}{c}{20}  & \multicolumn{2}{c}{80} \\
                                             & $^{18}$Ne & $^{18}$Ne   &  $^{18}$Ne   &  $^{6}$He    \\
\hline
SPS cycle, $T_C$, [s]                        & 3.6      &       6.0   &   14.4      &  16.8         \\
Injected [10$^{11}$/bunch]                    & 1.0      &      3.8    &   0.9      &  2.5           \\
Accumulated [10$^{12}$/bunch]                 & 1.3      &      3.0    &   0.8      & 1.4             \\
Accumulated [10$^{13}$/batch]                 & 2.5      &      6.1    &   6.7      & 11.3           \\
$\nu$-flux, $\phi$, [10$^{18}$/year]          & 0.4     & 2.0       & 1.1        & 3.5           \\
Nom. flux, $\phi_0$, [10$^{18}$/year]         & 1.1     & 2.9       & 1.1        & 2.9           \\
$\nu$-flux ratio, $\phi/\phi_0$               & 0.7     & 0.4       & 1.0        & 1.2           \\
\end{tabular}
\caption{Capturing and merging simulation results for the low-Q isotopes after injected bunch intensities on the 2$^{nd}$ row are assumed. 
  Neutrino fluxes are calculated using eq.~(\ref{eq:phi}).
}
\label{tab:ION-ACCUM-RESULT}
\end{table}


\section{Conclusions}
\label{conclusions}

A detailed study program for Beta Beam implementation at CERN has been presented in this report. 
There has been encouraging ion production progress towards required source rate, specifically 
the $^{18}$Ne production experiment will take place at CERN in November 2011. 
There are well established beam preparation studies, for example a state of the art studies 
for a 60~GHz ECR is ongoing. 
The new indication of large $\theta_{13}$ has opened many new possibilities for the Beta Beams.
The result of the impact-studies of one of them were presented here. We saw that the DR now can
have injected and store enough ions to reach required neutrino flux for the low-Q ion pair baseline. 



\begin{theacknowledgments}
I acknowledge the financial support of the European Community under the
European Commission Framework Programme 7 Design Study: EUROnu, Project
Number 212372. The EC is not liable for any use that may be made of the
information contained herein.

\end{theacknowledgments}

\bibliographystyle{aipproc}   
\bibliography{References}

\begin{thebibliography}{32}
\expandafter\ifx\csname natexlab\endcsname\relax\def\natexlab#1{#1}\fi
\providecommand{\enquote}[1]{``#1''}
\expandafter\ifx\csname url\endcsname\relax
  \def\url#1{\texttt{#1}}\fi
\expandafter\ifx\csname urlprefix\endcsname\relax\def\urlprefix{URL }\fi
\providecommand{\eprint}[2][]{\url{#2}}

\bibitem[Fukuda et~al.(1998)]{SK-OSCILLATION}
Y.~Fukuda, et~al., \emph{Phys. Rev. Lett.} \textbf{81}, 1562--1567,  (1998),
  \eprint{hep-ex/9807003}.

\bibitem[Schwetz et~al.(2011)]{Schwetz:2011zk}
T.~Schwetz, M.~Tortola, and J.~W.~F. Valle  (2011), \eprint{1108.1376}.

\bibitem[Abe et~al.(2011)]{Abe:2011sj}
K.~Abe, et~al.  (2011), \eprint{1106.2822}.

\bibitem[{The Commission of the European Communities}(2007)]{FP7-EURONU}
{The Commission of the European Communities}, \emph{{EUROnu - A High Intensity
  Neutrino Oscillation Facility in Europe}}, 2007,
  \eprint{FP7-INFRASTRUCTURES-2007-1}.

\bibitem[Zucchelli(2002)]{ZUCCHELLI-BETABEAM}
P.~Zucchelli, \emph{Phys. Lett.} \textbf{B532}, 166--172,  (2002).

\bibitem[Fernandez-Martinez(2010)]{FernandezMartinez:2009hb}
E.~Fernandez-Martinez, \emph{Nucl. Phys.} \textbf{B833}, 96--107,  (2010),
  \eprint{0912.3804}.

\bibitem[Kugler(1993)]{Kugler:1993vm}
E.~Kugler, \emph{Nucl.Instrum.Meth.} \textbf{B79}, 322--325,  (1993).

\bibitem[D'Elia et~al.(2009)]{DElia:2009xb}
A.~D'Elia, R.~Jones, and M.~Pasini  (2009), \eprint{0910.0409}.

\bibitem[Mendonca(2011)]{MENDONCA_NUFACT11}
T.~Mendonca, \emph{Opportunities for neutrino experiments at ISOLDE}, To be
  Published in AIP Conference Proceedings for NUFACT2011, 2011.

\bibitem[Rubbia et~al.(2006)]{RUBBIA-BEAMCOOLING}
C.~Rubbia, A.~Ferrari, Y.~Kadi, and V.~Vlachoudis, \emph{Nucl. Instrum. Meth.}
  \textbf{A568}, 475--487,  (2006), \eprint{hep-ph/0602032}.

\bibitem[Schaumann(????)]{Schaumann:2009zz}
M.~Schaumann CERN-THESIS-2009-128.

\bibitem[Mitrofanov(2011)]{SEMEN_NUFACT11}
S.~Mitrofanov, \emph{Collection of 8B and 8Li}, To be Published in AIP
  Conference Proceedings for NUFACT2011, 2011.

\bibitem[Kravchuk(2011)]{VLADIMIR_NUFACT11}
V.~Kravchuk, \emph{8B Production Measurements for the FP7 Beta Beam Design
  Study}, EUROnu WP4 Collaboration Note, 2011.

\bibitem[Benedetto(2010)]{Benedetto:2010zza}
E.~Benedetto p. MOPD069,  (2010).

\bibitem[Nolen(2011)]{NOLEN_NUFACT11}
J.~Nolen, \emph{Liquid targets for isotope production}, To be Published in AIP
  Conference Proceedings for NUFACT2011, 2011.

\bibitem[Stora(2011)]{STORA_NUFACT11}
T.~Stora, \emph{Ion Production for Beta-beams}, To be Published in AIP
  Conference Proceedings for NUFACT2011, 2011.

\bibitem[Lamy(2011)]{LAMY_NUFACT11}
T.~Lamy, \emph{60 GHz ECR source status}, To be Published in AIP Conference
  Proceedings for NUFACT2011, 2011.

\bibitem[Lachaize(2008)]{Lachaize:2008zz}
A.~Lachaize, \emph{PoS} \textbf{NUFACT08}, 087,  (2008).

\bibitem[Benedikt et~al.(2006)]{Benedikt:2006zk}
M.~Benedikt, A.~Fabich, M.~Kirk, C.~Omet, and P.~Spiller pp. 1696--1698,
  (2006).

\bibitem[Ruggiero(1995)]{Ruggio:1995}
F.~Ruggiero  (1995), cERN-SL/95-09 (AP), LHC Note 313.

\bibitem[Wildner et~al.(????)]{WILDNER-MATHEMATICA}
E.~Wildner, et~al. CERN-AB-2007-015.

\bibitem[Chanc\'{e} et~al.(2011)]{ANTOINE_NUFACT11}
A.~Chanc\'{e}, et~al., \emph{A new lattice for the beta-beam decay ring to
  enlarge the stability limit}, To be Published in AIP Conference Proceedings
  for NUFACT2011, 2011.

\bibitem[Chance and Hancock(????)]{HANCOCK-STACKING}
A.~Chance, and S.~Hancock Prepared for European Particle Accelerator Conference
  (EPAC 06), Edinburgh, Scotland, 26-30 Jun 2006.

\bibitem[Heinrich et~al.(2010)]{HEINRICH}
D.~Heinrich, C.~Hansen, and A.~Chance, \emph{Simulations of Bunch Merging in a
  Beta Beam Decay Ring}, To be Published in AIP Conference Proceedings for
  NUFACT2010, 2010.

\bibitem[{Rumolo, Giovanni et al}(????)]{headtail}
{Rumolo, Giovanni et al} CERN-SL-Note-2002-036-AP.

\bibitem[Chin(????)]{Chin:1988xj}
Y.~H. Chin CERN-LEP-TH/88-05.

\bibitem[Metral(2004)]{Metral:2004vi}
E.~Metral, \emph{{Overview of single-beam coherent instabilities in circular
  accelerators}}, Prepared for 1st CARE-HHH-APD Workshop on Beam Dynamics in
  Future Hadron Colliders and Rapidly Cycling High-Intensity Synchrotrons,
  Geneva, Switzerland, 8-11 Nov 2004, 2004.

\bibitem[Chin(1998)]{Chin:8thJoint}
Y.~H. Chin, \emph{{Impedance and Wakefields}}, Proceedings of the 8th Joint
  School on Accelerator Physics, 1998.

\bibitem[Fischer et~al.(????)]{Fischer:2008zzc}
W.~Fischer, et~al. EPAC'08, 11th European Particle Accelerator Conference, 23-
  27 June 2008, Genoa, Italy.

\bibitem[Rumolo(2011)]{rumolo_private}
G.~Rumolo  (2011), {Private discussions}.

\bibitem[Benedikt et~al.(2011)]{FP6-FINAL}
M.~Benedikt, et~al., \emph{Eur. Phys. J.} \textbf{A47}, 24,  (2011).

\bibitem[Chance(2011)]{ANTOINE_IPAC11}
A.~Chance, \emph{A New Lattice for the Beta-Beam Decay Ring to Mitigate the
  Head Tail Effects}, To be Published in the JACoW Conference Proceedings for
  IPAC2011, 2011.

\end{thebibliography}

\end{document}